\documentclass{aa}  

\usepackage{graphicx}
\usepackage{txfonts}
\usepackage{mathrsfs}
\usepackage[colorlinks,allcolors=blue]{hyperref}
\usepackage{breakurl}
\usepackage{subfigure}
\usepackage{stackengine}
\usepackage{amsmath}

\newcommand{\inte}{INTEGRAL}

\newcommand {\beq}{\begin {eqnarray}}
\newcommand {\eeq}{\end {eqnarray}}
\renewcommand{\d}{{\rm d}}

\begin{document}

\title{Searching for redshifted 2.2~MeV neutron-capture lines from accreting neutron stars: Theoretical X-ray luminosity requirements and INTEGRAL/SPI observations}
\titlerunning{Redshifted 2.2~MeV neutron-capture lines from accreting neutron stars}

\author{L.~Ducci
\inst{1,2,3}
\and
A.~Santangelo
\inst{1}
\and
S.~Tsygankov
\inst{4}
\and
A.~Mushtukov
\inst{5}
\and
C.~Ferrigno
\inst{2,3}
}

\institute{Institut f\"ur Astronomie und Astrophysik, Kepler Center for Astro and Particle Physics, University of Tuebingen, Sand 1, 72076 T\"ubingen, Germany\\
\email{ducci@astro.uni-tuebingen.de}
\and      
ISDC Data Center for Astrophysics, Universit\'e de Gen\`eve, 16 chemin d'\'Ecogia, 1290 Versoix, Switzerland
\and
INAF -- Osservatorio Astronomico di Brera, via Bianchi 46, 23807 Merate (LC), Italy
\and
Department of Physics and Astronomy, University of Turku, 20014 Turku, Finland
\and
Astrophysics, Department of Physics, University of Oxford, Denys Wilkinson Building, Keble Road, Oxford OX1 3RH, UK
}

   \abstract
       {Accreting neutron stars (NSs) are expected to emit a redshifted $2.2$~MeV line due to the capture of neutrons produced through the spallation processes of $^4$He and heavier ions in their atmospheres. Detecting this emission would offer an independent method for constraining the equation of state of NSs and provide valuable insights into nuclear reactions occurring in extreme gravitational and magnetic environments.
Typically, a higher mass accretion rate is expected to result in a higher $2.2\,{\rm MeV}$ line intensity. However, when the mass accretion rate approaches the critical threshold, the accretion flow is decelerated by the radiative force, leading to a less efficient production of free neutrons and a corresponding drop in the flux of the spectral line.
This makes the brightest X-ray pulsars unsuitable candidates for $\gamma$-ray line detection.
In this work, we present a theoretical framework for predicting the optimal X-ray luminosity required to detect a redshifted 2.2 MeV line in a strongly magnetized NS.
As the INTErnational Gamma-Ray Astrophysics Laboratory (\inte) mission nears its conclusion, we have undertaken a thorough investigation of the SPectrometer on board \inte\ (SPI) data of this line in 
a representative sample of accreting NSs.
No redshifted 2.2~MeV line was detected. For each spectrum, we have determined the $3\sigma$ upper limits of the line intensity, 
assuming different values of the line width.
Although the current upper limits are still significantly above the expected line intensity, 
they offer valuable information for designing future gamma-ray telescopes aimed at bridging the observational MeV gap.
Our findings suggest that advancing our understanding of the emission mechanism of the $2.2$~MeV line, as well as the accretion flow responsible for it, will require a substantial increase in sensitivity from future MeV missions.
For example, for a bright X-ray binary such as Sco~X$-$1, we would need at least a 3$\sigma$ line point source sensitivity of $\approx 10^{-6}$~ph~cm$^{-2}$~s$^{-1}$, that is, about two orders of magnitude better than that currently achieved.
       }

  \keywords{accretion -- stars: neutron gamma-rays: stars -- X-rays: binaries -- nuclear reactions}

   \maketitle

\section{Introduction}

A seminal paper by \citet{Shvartsman70} showed that accreting neutron stars (NSs) are not merely X-ray emitters; the accreting matter can also produce spectral features in the MeV band.
These spectral features are possible because, as matter falls onto a NS, its kinetic energy can exceed the binding energy of nucleons within nuclei, leading to their disintegration upon atmospheric entry.
A fraction of the accreting ions can thus produce nuclear reactions in the external layer of the NS, yielding $\gamma-$ray line emission.
One of the $\gamma-$ray production channels produces the emission of $2.2$~MeV photons as a result of the capture of neutrons by protons.
This idea was further developed in subsequent studies by \citet{Reina74} and \citet{Brecher80}. 
Studies by \citet{Bildsten92,Bildsten93} continued to build on these works, advancing a more robust photon production mechanism, capable of determining the expected intensity of the $2.2$~MeV emission line. The main aspects of this model are briefly described below.

The accreting nuclei (here, we are especially interested in $^4$He) slow down due to repeated Coulomb collisions with electrons in the NS atmosphere. Then, their collision with atmospheric protons results in the destruction of a significant fraction of the nuclei, with the consequent release of neutrons and $^3$He.
The neutrons liberated in the spallation reactions continue to scatter elastically with protons, eventually thermalizing and drifting downwards under the influence of gravity.
They can then be captured by protons, resulting in the emission of $2.2$~MeV photons, or undergo charge exchange with $^3$He.
\citet{Bildsten93} pointed out that the production of $2.2$~MeV photons is primarily from neutrons in excess of the $^3$He that is produced, as most $^3$He will 
absorb a neutron, leading to the production of a proton and $^3$H. The capture of the proton by a deuterium leads to a $\gamma-$ray with an energy of $\sim 5.49$~MeV, while the capture of a proton by $^3$H produces a $\gamma-$ray with an energy of $\sim 19.81$~MeV \citep[see ][ and references therein]{Bildsten93}.
Due to the high Compton optical depths within the NS atmosphere, only a fraction of the $2.2$~MeV $\gamma-$rays manage to escape without scattering.

In their theoretical model, \citet{Bildsten93} considered a scenario with a moderate magnetic field, where matter accretes radially onto the NS. In the presence of stronger magnetic fields, approximately $\sim 10^{12}$~G in accreting pulsars of high-mass X-ray binaries (see e.g. \citealt{2022arXiv220414185M}), the escape of these line photons from the NS is determined by the potential attenuation in their magnetosphere, whose leading process is the magnetic pair production.
\citet{Caliskan09} provided a rough estimate for transparency to $2.2$~MeV photons due to pair creation: for $\gamma-$ray photons to be observable, their origin must be at polar magnetic co-latitudes below approximately 20 degrees and they must be beamed within 15-20 degrees relative to the local surface magnetic field vector. These conditions restrict the likelihood of observing a $2.2$~MeV emission line to small, yet non-negligible, probabilities.
Since this line originates from the radially thin atmosphere of the NS, its detection would enable determination of the gravitational redshift and thus constrain the nuclear equation of state at high matter densities. The $2.2$~MeV emission line produced in NSs is also expected to be broadened, due to gravitational and relativistic effects \citep{Ozel03}.
In summary, the $2.2$~MeV emission line has the potential to be a crucial spectroscopic tool in the study of high-energy environments at the surfaces and in the interiors of NSs. Detecting this line could provide valuable insights into nuclear reactions occurring in accretion flows within environments characterized by strong gravitational and magnetic fields.

While the redshifted $2.2$~MeV line from the surface of the accreting NS is the focus of our work here, we briefly mention a few other potential origins of the $2.2$~MeV line emission, in the interest of thoroughness.
These include lines originating from the hot inner region of accretion discs around black holes \citep{Aharonian84}, narrow lines resulting from neutron capture on the surface of the donor star in the binary system \citep{Jean01,Guessoum02, Guessoum04}, and emission of the $\sim 2.22$~MeV line due to very high energy protons impacting the surface of the donor star \citep{Vestrand89}.

Studies have been carried out to search for narrow, unredshifted $2.2$~MeV line emission.
\citet{Harris91} placed initial constraints on $2.2$~MeV emissions from the low-mass X-ray binary Scorpius~X-1 (Sco X-1) at $10^{-4}$~ph~cm$^{-2}$~s$^{-1}$ using data from the Solar Maximum Mission. Subsequently, a comprehensive all-sky survey for $2.2$~MeV emissions was carried out with the Compton Telescope (COMPTEL) aboard the Compton Gamma Ray Observatory. This survey provided tighter constraints on several X-ray binaries (XRBs) including an upper limit on the emission from Sco~X-1 of $2.5 \times 10^{-5}$~ph~cm$^{-2}$~s$^{-1}$ as reported by \citet{McConnell97}.
The study by \citet{Teegarden06} reported less stringent upper limits on emission lines from various XRBs, utilizing the initial year of data from \inte/SPI.
The redshifted $2.2$~MeV line from A~0535+26 has been the subject of studies by \citet{Boggs06} and \citet{Caliskan09}.
This target was selected because the pulsar in A~0535+26 is relatively slow ($\sim100$~s; \citealt{Rosenberg75}), and hence the rotational broadening should be negligible.
\citet{Boggs06} analysed data collected by the Reuven Ramaty High Energy Solar Spectroscopic Imager (RHESSI) during a giant outburst in 2005, while \citet{Caliskan09} examined \inte/SPI data from a fainter outburst that occurred the same year, three months after the major event. Neither study detected the redshifted $2.2$~MeV line; both were only able to establish width-dependent upper limits ranging from $(2-11) \times 10^{-4}$ ph cm$^{-2}$ s$^{-1}$.

Given the importance of detecting the redshifted $2.2$~MeV neutron-capture line emission, in Sect. \ref{sec: model} we present a theoretical framework that provides, for the first time, the optimal X-ray luminosity for the production of 2.2~MeV photons from a highly magnetized accreting NS.
In Sect. \ref{sect. data analysis} we present the most comprehensive analysis of \inte/SPI data on the subject to date. Our analysis consists of searching for an emission line in the $\sim1-2.2$~MeV energy band using approximately 20 years of SPI data from observations of a representative sample of accreting NSs. The results are presented in Sect. \ref{sect. results} and discussed in Sect. \ref{sec: discussion}.

\section{Critical luminosity effects on 2.2~MeV line emission in highly magnetized NSs}
\label{sec: model}

The theory of the emission of the $2.2$~MeV line from accreting NSs (see the references in the Introduction) was developed mainly for stars with a weak magnetic field. 
An extremely strong magnetic field, $\gtrsim 10^{11}\,{\rm G}$, usually observed in X-ray pulsars (XRPs), should result in significant changes to the model. 
Detailed modification of the existing theory is a topic for a separate publication. 
Here, however, we do discuss one of the most important effects, which relates to the existence of the critical luminosity \citep[][]{1976MNRAS.175..395B,2012A&A...544A.123B,2015MNRAS.447.1847M}, when the accretion flow can be decelerated above the stellar surface as a result of its interaction with X-ray photons.

\subsection{Simplified model for non-magnetic NSs}
\label{sec:non-magnetic}

Since the appearance of the $2.2\,{\rm MeV}$ $\gamma$-ray photons is a result of neutron capture by protons following the destruction of helium atoms,  
the $\gamma$-ray line flux is expected to be proportional to the accretion rate of $^4$He nuclei onto a NS surface \citep{Bildsten93}.
As a result, both X-ray flux, $F_{\rm X}$, and the $\gamma-$ray line flux, $F_{\rm n,\,2.2\rm MeV}$, are scaled linearly with the mass accretion rate.
The flux of the $2.2\,{\rm MeV}$ photon number can be estimated as
\beq \label{eq:N_flux}
F_{\rm n,\,2.2 MeV} \equiv \frac{\d N_{2.2 \rm MeV}}{\d t \d S} = Q\left(\frac{Y}{4}\right)
\frac{F_{\rm X} R_{\rm NS}}{GM_{\rm NS}m_{\rm p}},
\eeq 
where
$Y$ is the mass fraction of $^4$He atoms in the accretion flow, 
$Q$ is the number of unscattered $2.2\,{\rm MeV}$ photons escaping the NS atmosphere per one atom of accreted helium, and the total X-ray energy flux,
\beq \label{eq:F_x}
  F_{\rm X} &    =   & \frac{1}{4 \pi d^2} \label{Fx}\dot{N}_{\rm b} E_{\rm i},
\eeq 
is determined by the accretion rate of baryons $\dot{N}_{\rm b}$, by
the kinetic energy of a baryon entering the NS atmosphere,
\beq \label{eq:E_i}
  E_{\rm i}       & \approx & \frac{G M_{\rm NS} m_{\rm p}}{R_{\rm NS}}\sim 0.22\,m_{\rm p} c^2,
\eeq
and by the distance to the source, $d$.
Estimations (\ref{eq:N_flux}) and (\ref{eq:F_x}) assume that the beam patterns in X-ray and the $\gamma$-ray line are similar and isotropic. 
\footnote{
The anisotropy of X-ray and $\gamma-$ray emission from the NS surface causes a difference between the isotropic and observed phase-averaged flux, but the difference is expected to be within a factor of two (see e.g. \citealt{2024MNRAS.527.5374M}).} 

%
Assuming a linear relation between the X-ray energy flux and the flux in the $\gamma$-ray line and $Q=0.2$ in (\ref{eq:N_flux}),
\citet{Bildsten93} estimated an upper limit to the expected 2.2~MeV $\gamma-$ray line photon flux 
from an accreting NS:
\beq \label{eq:bildsten} 
  F_{\rm n,\,2.2 MeV} &<& 1.4 \times 10^{-5} \mbox{ph~cm}^{-2}\mbox{s}^{-1} \left( \frac{Y}{0.25} \right) \left( \frac{162 \mbox{MeV}}{E_{\rm i}} \right) \nonumber \\
 &&\times \left( \frac{F_{\rm X}}{3\times 10^{-7} \mbox{erg~cm}^{-2}\mbox{s}^{-1}} \right) \mbox{.}
\eeq
However, the upper limit (\ref{eq:bildsten}) can be significantly higher than the real flux in the $2.2$\, MeV line, because $Q=0.2$ does not take several factors into account, namely, that not all of the accreted neutrons are liberated from their nuclei,
that neutrons tend to thermalize at large depths in the atmosphere, and that a large fraction of neutrons are absorbed by $^3$He nuclei.
\citet{Bildsten93} provided more accurate estimations for $Q$ at different infall energies of the nucleus $E_{\rm i}$ (see Table 4 in \citealt{Bildsten93}).
For the kinetic energy $E_{\rm i}\approx 162$~MeV, corresponding to the free-fall velocity at $M_{\rm NS}=1.4$M$_\odot$ and $R_{\rm NS}=12$~km NS surface, $Q\approx 10^{-2}$.
Thus, for convenience, Eq.\,\ref{eq:bildsten} can be re-written to show the expected $\gamma-$ray line flux:
\beq\label{eq bildsten ul2}
  F_{\rm n,\,2.2 MeV} &\sim& 1.2 \times 10^{-6} \mbox{ph~cm}^{-2}\mbox{s}^{-1} \left( \frac{Q}{10^{-2}} \right) \left( \frac{Y}{0.25} \right) \left( \frac{162 \mbox{MeV}}{E_{\rm i}} \right) \nonumber \\
 &&\times \left( \frac{F_{\rm X}}{3\times 10^{-7} \mbox{erg~cm}^{-2}\mbox{s}^{-1}} \right)  \mbox{\ .}
\eeq 

\subsection{Optimal luminosity of an XRP for the emission of 2.2~MeV photons}
\label{sec:optimal}

The efficiency of $2.2$\,MeV photon production per one $^4$He atom accreted onto the NS surface is dependent on the kinetic energy of the atom and decreases as the kinetic energy becomes smaller.
The number of unscattered $2.2\,{\rm MeV}$ photons escaping the NS atmosphere per one atom of accreted helium can be roughly approximated
\footnote{
The accuracy of approximation is $10$ per cent within the kinetic energy interval $E_{\rm i}\in [50\,{\rm MeV};250\,{\rm MeV}]$.
}
as (see Table\,4 in \citealt{Bildsten93})
\beq\label{eq:Q}
Q\approx 0.04\,\exp\left[ -\frac{300}{E_{\rm i, MeV}^{1.1}} \right],
\eeq 
where $E_{\rm i, MeV}\equiv E_{\rm i}/1\,{\rm MeV}$.
The kinetic energy of particles entering a stellar atmosphere depends on the star's mass and radius (see Eq. \ref{eq:E_i}), but
at high mass accretion rates, the luminosity and radiative force are large enough to decelerate accretion flow above the stellar surface \citep{1976MNRAS.175..395B}, which reduces the kinetic energy. 
At the critical luminosity $L_{\rm crit}\sim 10^{37}\,{\rm erg\,s^{-1}}$, which is expected to be dependent on the magnetic field strength at the NS surface \citep{2015MNRAS.447.1847M}, the flow is completely decelerated due to its interaction with X-ray photons, that is, $E_{\rm i}=0$.
At luminosities below the critical value, the velocity of accretion flow entering the atmosphere can be approximated as 
\beq \label{eq:v}
v \approx v_{\rm ff}\left(1 - \frac{L}{L_{\rm crit}}\right)^{1/2}, 
\eeq 
where $v_{\rm ff}$ is the free-fall velocity (see Sect.\,2 in \citealt{2015MNRAS.454.2714M} and \citealt{2023AstL...49..583M} for the advanced calculations). 
%
%
   \begin{figure}
   \centering
   \includegraphics[width=\columnwidth]{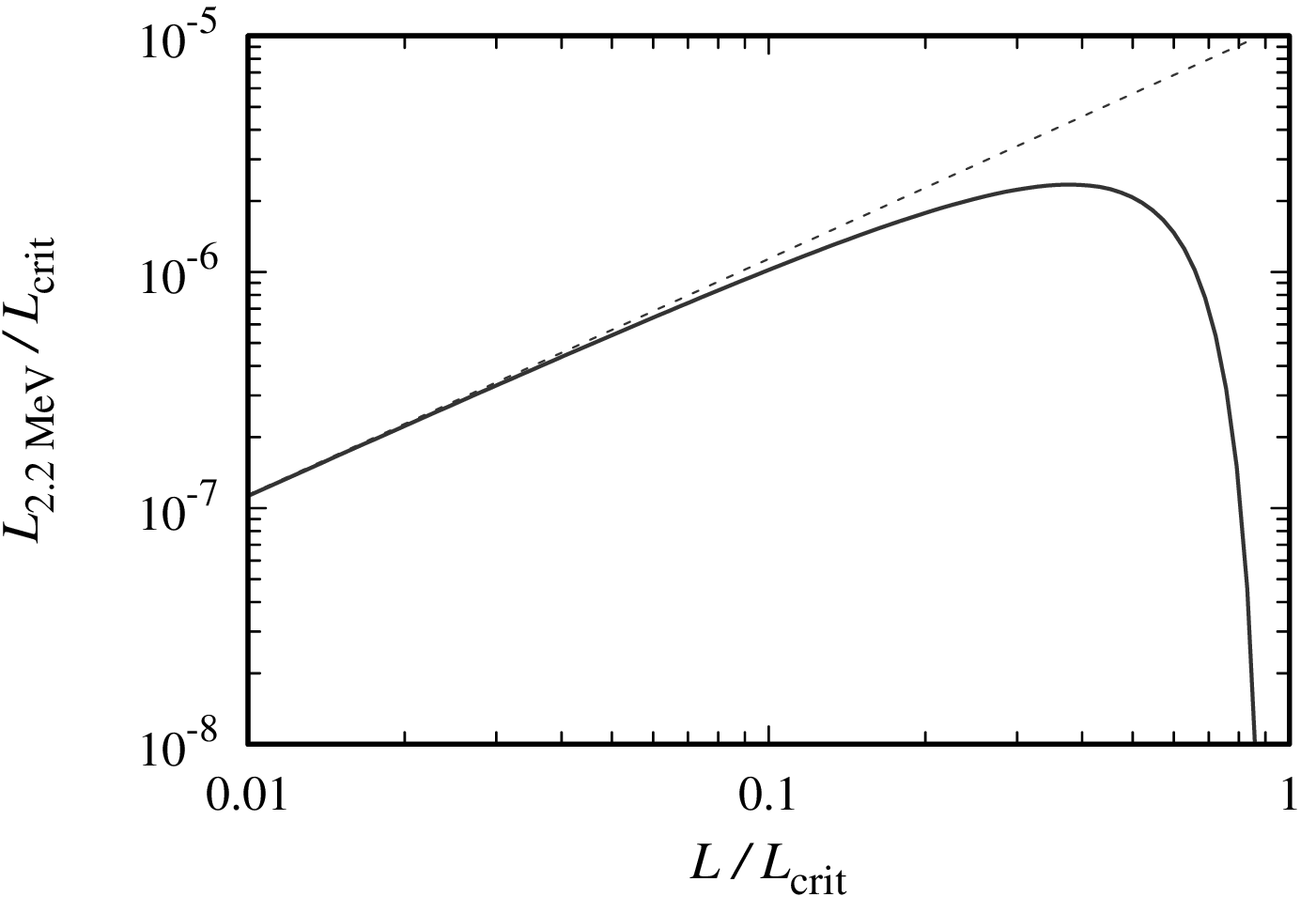}
   \caption{
   Expected luminosity of the 2.2 MeV line, $L_{2.2\,\rm MeV}$, as a function of the total accretion luminosity $L$ of an XRP.
   Both axes are scaled by the critical accretion luminosity $L_{\rm crit}$.
   At low mass accretion rates and luminosity, the flux in the 2.2 MeV line is proportional to the total accretion luminosity.
   At high mass accretion rates, however, the radiative force decelerates the accretion flow above the NS surface, which results in a sharp drop in luminosity in the 2.2 MeV line.}
   \label{fig:L_2.2}
   \end{figure}
%
%
%
Therefore, a naive linear correlation between the X-ray flux and the flux in the 2.2~MeV line given by Eq.~(\ref{eq bildsten ul2}) is not valid for XRPs near or above the critical luminosity.
Combining approximations (\ref{eq:N_flux}), (\ref{eq:Q}), and (\ref{eq:v}), we can estimate the luminosity and flux in the 2.2\,MeV line for an XRP (see Fig.\,\ref{fig:L_2.2}).
At low mass accretion rates, the luminosity (and flux) in the 2.2\,MeV line is expected to be linearly proportional to the total accretion luminosity, but drops very quickly as soon as the source reaches its critical luminosity.
Therefore, the maximal luminosity of the 2.2\,MeV line is achieved at $L\sim 0.5 L_{\rm crit}$ and is expected to be 
\beq\label{eq:L_2.2_max}
L_{2.2\,{\rm MeV}}^{\rm (max)}\sim 5\times 10^{-6} L_{\rm crit}.
\eeq 
For a typical XRP with a magnetic field of the order of $few\times10^{12}$~G and corresponding critical luminosity of around $10^{37}\,{\rm erg\,s^{-1}}$, the optimal X-ray luminosity for detecting the redshifted 2.2~MeV line is $\sim5\times10^{36}\,{\rm erg\,s^{-1}}$.
The main process determining accretion flow deceleration in XRPs - namely, Compton scattering - is strongly affected by the magnetic field \citep{1986ApJ...309..362D}.
As a result, the critical luminosity is anticipated to be highly dependent on the surface magnetic field strength of a NS, following the relation $L_{\rm crit}\propto B_0^2$ for $B_0\gtrsim 4\times 10^{12}\,{\rm G}$.
The primary uncertainties in determining the critical luminosity, given a known surface field strength, stem from the geometry of the accretion channel at the NS surface (it can be affected by global geometry of the NS magnetic field and the dipole magnetic field inclination with respect to the plane of the accretion disc) and the polarization composition of the X-rays emitted by the atmosphere of a NS (see Fig.\,5ce in \citealt{2015MNRAS.447.1847M}).

\begin{table}
\begin{center}
\caption{Time intervals of the observations analysed in this work.}
\vspace{-0.3cm}
\label{table:log1}
\resizebox{\columnwidth}{!}{
\begin{tabular}{lcc}
\hline
\hline
\noalign{\smallskip}
Source name         & time interval &  \inte\ revolution$^a$   \\
\noalign{\smallskip}
\hline
\noalign{\smallskip}
A~0535+26           &  17-18 August 2009 & 836 \\
                    &  17 March 2011     & 1028 \\
                    & 4-9 September 2019 & 2133, 2134\\
                    & 6-10 June 2020     & 2214, 2215 \\
                    & 14-15 April 2020   & 2217 \\
                    & 10-13 October 2021 & 2421, 2422\\
\noalign{\smallskip}
\hline
\noalign{\smallskip}
GX 304$-$1          & 17-18 January 2012 & 1131\\
                    & 4 February 2012    & 1137\\
                    & 7 February 2012    & 1138\\
\noalign{\smallskip}
\hline
\noalign{\smallskip}
Vela X$-$1          &  12 June 2003 - 1 July 2022 & 81-2520 \\
\noalign{\smallskip}
\hline
\noalign{\smallskip}
X~Persei            & 3 August 2004 - 7 August 2019 & 220-2122 \\
\noalign{\smallskip}
\hline
\noalign{\smallskip}
Sco X$-$1           & 27 March 2003 - 1 February 2023 & 55-2601\\
\noalign{\smallskip}
\hline 
\end{tabular}
}
\end{center}
 {\small Notes. \\
 \noindent $^a$ The numbers of \inte\ revolutions around the Earth (starting from 0 at launch) are commonly used to refer to the corresponding observations.
  }
\end{table}

\section{Data analysis}
\label{sect. data analysis}

We searched for the presence of a redshifted 2.2~MeV line in SPI observations of accreting pulsars. SPI \citep{Vedrenne03} operates in the 20~keV$-$8~MeV energy range, with an energy resolution of $\sim 2.8$~keV at 1.7~MeV (see e.g. \citealt{Roques03, Diehl18}).
We performed the data reduction using the SPI Data Analysis Interface ({\tt spidai}) software provided by the SPI team at the IRAP Toulouse\footnote{{\tt spidai} is available at \url{https://sigma2.irap.omp.eu/integral/spidai} by requesting a personal account.}.
Spectra are derived using a sky model-fitting procedure, with a background determined using a set of empty fields observations, called 'flat fields', roughly one per six months. The flat fields are pre-defined and are part of the {\tt spidai} configuration.
The {\tt spidai} sky model fitting algorithm adopts the minimization of the $\chi^2$ statistics. Its application to Poisson-distributed count noise - as, in our case, it is for narrow bins at high energies - might underestimate the total number of counts \citep{Mighell99}. 
Therefore, the outcomes of the sky model setup were interpreted with due caution. 
In particular, 
we conducted an intermediate test to setup the sky model using fewer bins (five, on a logarithmic scale) from 20~keV to 2.2~MeV, which allowed for a higher number of counts per bin. These bins also include the energy range in which the sources are bright ($\lesssim100$~keV), which helps determine the sources contributing to the total flux measured by SPI, thereby improving the input sky model. The sky model solutions, built on the basis of the intermediate step but now with many narrow bins in the range 20~keV$-$2.2~MeV, were then checked to detect any problems, such as anomalous high values of $\chi^2$ in individual energy and pointing\footnote{\inte\ pointings, called science windows, typically last 1-3~ks. During each pointing, the satellite stares at a specific position.} channels (see e.g. \citealt{Jourdain09} for a detailed description of the method). Pointings exhibiting unusually high background activity were excluded.
We analysed public data where the targets selected for this work were within 12$^\circ$ from the centre of the SPI field of view.

\begin{table*}
\begin{center}
\caption{Accreting NSs analysed in this study, along with their respective net exposure times for the SPI data, and the highest $3\sigma$ upper limits in the ranges 1$-$2.2~MeV (for a redshifted 2.2~MeV line) and 200-800~keV (for a 511~keV line), assuming different values of FWHM.}
\vspace{-0.3cm}
\label{Table sources}
\resizebox{\columnwidth+\columnwidth}{!}{
\begin{tabular}{lcccccccccccc}
\hline
\hline
\noalign{\smallskip}
Source name         & net exposure &  $L_{\rm x}$$^a$ & distance$^b$         & \multicolumn{5}{c}{$3\sigma$ u.l. 2.2~MeV line}  &  \multicolumn{4}{c}{$3\sigma$ u.l. 511~keV line} \\
\noalign{\smallskip}
                    &   (ks)       &  (erg~s$^{-1}$)   &  (kpc)              & \multicolumn{5}{c}{$10^{-4}$ph~cm$^{-2}$~s$^{-1}$}  &   \multicolumn{4}{c}{$10^{-3}$ph~cm$^{-2}$~s$^{-1}$}     \\
                    &        &                   &                     &FWHM:&    10   &    20   &    40   &    100    &FWHM:&    10   &    20     &     40    \\
                    &        &                   &                     &(keV)&         &         &         &           &(keV)&         &           &           \\
\noalign{\smallskip}
\hline
\noalign{\smallskip}
A~0535+26            & 1631         &  $3.1\times 10^{37}$   & $1.8\pm0.1$   &           &   7.5   &  9.5    &   11     & 16      &  &   1.5  &   1.1     &     1.6    \\
\noalign{\smallskip}
GX~304$-$1          & 381          &  $2.6\times 10^{36}$   & $1.86\pm0.04$  &            &   18   &  15      &  22     &   36    &   &  2.5  &   2.5      &    2.8     \\
\noalign{\smallskip}
Vela X-1          & 5345         & $2.8\times 10^{36}$    & $1.96\pm0.06$    &            &   2.4   &  3.0    &   3.8   &   7.2   &  &   0.7   &   0.7    &   0.8       \\
\noalign{\smallskip}
X~Persei          & 3050           & $6\times 10^{34}$    & $0.60\pm0.02$    &            &   4.4   &  5.2    &   8.2   &   9.8    &  &  0.4   &   0.5     &  0.8          \\
\noalign{\smallskip}
Sco X-1           & 4495         & $1.6\times 10^{38}$    & $2.1\pm0.1$      &            &   2.8   &  3.6     &  5.3   &   9.6    &  &  0.3    &   0.3    &  0.4            \\
\noalign{\smallskip}
\hline 
\end{tabular}
}
\end{center}
 {\small Notes. \\
 \noindent $^a$ X-ray luminosity, in the energy range 0.1--100~keV.\\
 \noindent $^b$ distances from \citet{Bailer21}.\\
 }
\end{table*}

We have listed the sources considered in this work in Table \ref{Table sources}.
These binary systems are located in uncrowded regions of the sky, which ensures a relatively simple sky model and therefore facilitates the possible detection of a faint spectral emission line.
Four of these XRBs host a slow pulsating NS (spin period in the range of $\sim 100-800$~s). We selected them because they are expected to produce a relatively narrow emission line \citep{Ozel03}.
On the other hand, the NSs in these four XRBs are highly magnetized ($B\gtrsim 10^{12}$~G). Therefore, based on the results in Sect. \ref{sec:optimal}, we considered only SPI observations where these sources exhibit $L_{\rm x} \lesssim 0.5L_{\rm crit}$ (see Fig. \ref{fig:L_2.2}).
These observations are summarized in Table \ref{table:log1}.

A~0535+26 and GX~304$-$1 are transient Be/XRBs showing sporadic outbursts which can reach high luminosities ($L_{\rm peak} >  10^{37}$~erg~s$^{-1}$). For these,
we considered the rise and decay parts of the outbursts, where the X-ray luminosity is optimal for detecting the redshifted 2.2~MeV line. 
Vela~X-1 is a persistent high-mass XRB, with an orbital period of $\sim8.96$~d and a long eclipse ($\Delta \phi \approx 0.1697$). For this source, we considered all of the available public SPI data, excluding the eclipses according to the ephemeris reported in \citet{Falanga15}. 
X~Persei (X~Per) is a persistent Be/XRB with a pulsar ($P_{\rm spin}\approx 835$~s, \citealt{White76}) accreting at a low rate ($L_{\rm x}\approx 6\times 10^{34}$~erg~s$^{-1}$, assuming $d=600$~pc; \citealt{Reig99,Bailer21}). For this source we considered all of the available public SPI data.
The fifth target of our study, Sco~X-1, is a low-mass XRB with a NS accreting at a high rate \citep[see e.g.][]{Vrtilek91}.
It has been considered in other previous observational and theoretical works \citep{McConnell97, Harris91, Bildsten93} on the 2.2~MeV emission line.
Sco X-1 is probably a millisecond pulsar (although the pulsation has not yet been discovered; see \citealt{Galaudage22}), which would lead to a very broad
emission line \citep{Ozel03}.
In addition, it likely has a low magnetic field, allowing us to test the simplified model for the 2.2~MeV emission for non-magnetic NSs proposed by \citet{Bildsten93} (Sect. \ref{sec:non-magnetic}).
In the Appendix, we report the results of a complementary data analysis, using SPI observations of highly magnetized NSs in Be/XRBs during outbursts, that is, when the X-ray luminosity exceeds $L_{\rm crit}$, for which we expect highly inefficient neutron capture.

We extracted all of the spectra in the energy range of 20~keV to 2~MeV, which is the range enabled by the {\tt spidai} software, using 500 bins defined on a linear scale.
In the 400$-$2200~keV energy range ($\sim 650-2200$~keV until revolution 1170, in May 2012), the SPI data
exhibit spectral artefacts due to high-energy particles that saturate the main electronic chain,
known as the analogue front-end electronics (AFEE), resulting in false triggers.
To circumvent this issue, we used the output from a second electronic
chain, the pulse shape discriminator (PSD), which operates independently of the AFEE.
The PSD sub-system has an efficiency of $\sim 85$\% within its energy domain, attributed to a relatively high dead time.
To account for this efficiency, we applied a correction factor of 0.85 to the data and error bars
for the energy band of the spectra acquired using the PSD (for more information, see \citealt{Jourdain09, Roques19}).

For the spectral analysis, we used {\tt xspec}\footnote{{\tt XSPEC} version 12.13.1d \citep{Arnaud96}.} through {\tt PyXspec}\footnote{\url{https://heasarc.gsfc.nasa.gov/xanadu/xspec/python/html/index.html}}.

   \begin{figure}
   \centering
   \includegraphics[width=\columnwidth]{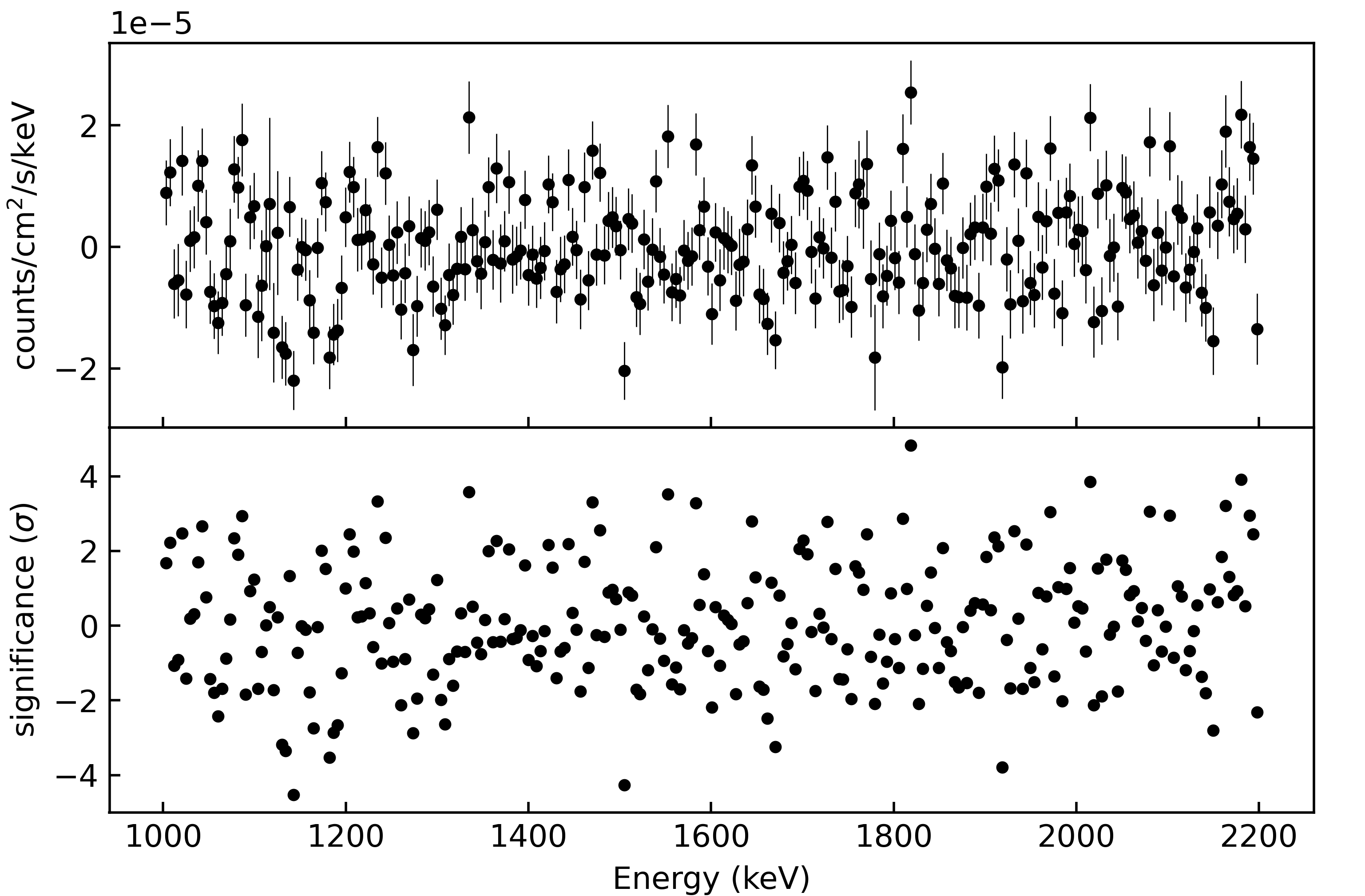}
   \caption{SPI spectrum of A~0535+26, in the energy range 1$-$2.2~MeV. \emph{Upper panel:} Background-subtracted spectrum of A~0535+26. \emph{Lower panel:} Significance of the residuals as a function of energy. Uncertainties in these panels do not include systematic errors.}
   \label{fig: example spectrum}
   \end{figure}

\section{Results} \label{sect. results}

\begin{figure}
   \centering
   \includegraphics[width=\columnwidth]{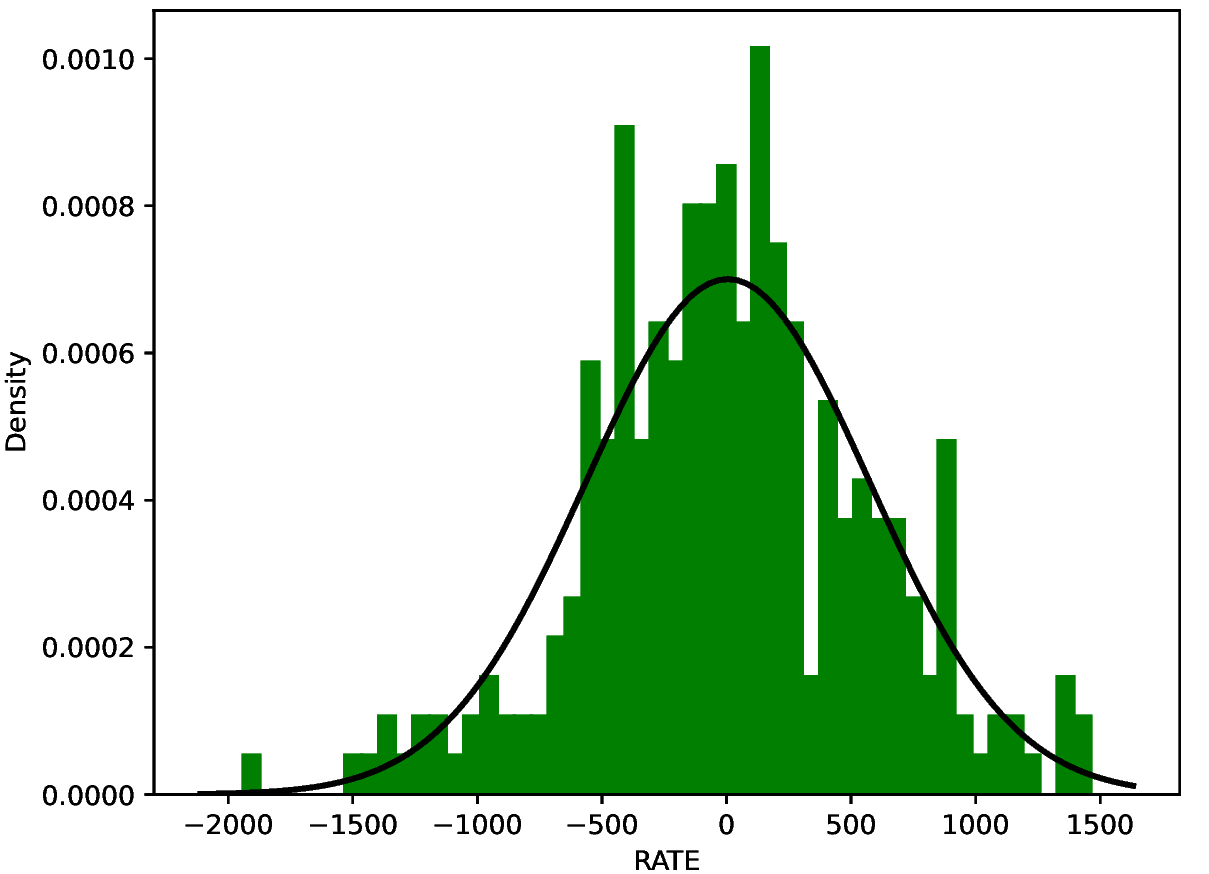}
   \caption{Green histogram: Distribution of the rate of the spectrum of A~0535+26 ($1-2.2$~MeV). Black line: Best fit with a Gaussian function.}
   \label{fig:gauss}
   \end{figure}

The {\tt spidai} data reduction software produces background-subtracted spectra in which, for each channel, rate and its error are provided in counts per second. 
The top panel of Fig. \ref{fig: example spectrum} shows an example (A~0535+26) of one of these spectra.
The spectra do not show any obvious hints indicating the presence of an emission line with a full width half maximum (FWHM) $\geq 10$~keV. 
In all of the spectra, the rate shows a symmetric distribution around zero (see, for example, the histogram of the rate of the $1-2.2$~MeV spectrum of A~0535+26 in  Fig. \ref{fig:gauss}). This is an effect that occurs when the background is subtracted, which is done when
the contribution from the source is low or absent (see e.g. \citealt{Siemiginowska11}). 
The lower panel of Fig. \ref{fig: example spectrum} shows that the fluctuations around zero are much higher than those expected from the statistical errors, which are probably 
underestimated.
We calculated systematic errors by empirically adding them in quadrature to the statistical errors until we obtained a reduced $\chi^2\sim 1$.
The expected energy and profile of the line depend on gravitational and relativistic effects, although it is not possible to determine in advance exactly how much it will be redshifted and broadened \citep[see ][ for a thorough discussion]{Ozel03}.
Given these uncertainties, to estimate the 3$\sigma$ upper limit of the line flux, we applied a method similar to that adopted by \citet{Boggs06} and \citet{Caliskan09}.
We fit each spectrum with a Gaussian over a grid of evenly spaced line energies and FWHMs.
The energy varies from 1.0~MeV to 2.2~MeV, with steps of 5~keV, and the FWHMs are fixed to 10, 20, 40, and 100~keV.
Table \ref{Table sources} shows, for each FWHM, the highest upper-limit values derived in the range $1-2.2$~MeV.
As an additional result, we report in Appendix \ref{app. B} the upper limits of the fluxes we obtained in the 200-800~keV and 1000-2200~keV bands by fitting the spectra with a power law ({\tt pegpwrlw} in {\tt xspec}).

\section{Discussion} \label{sec: discussion}

\subsection{SPI upper limits vs expectations}

\begin{figure}
   \centering
   \includegraphics[width=\columnwidth]{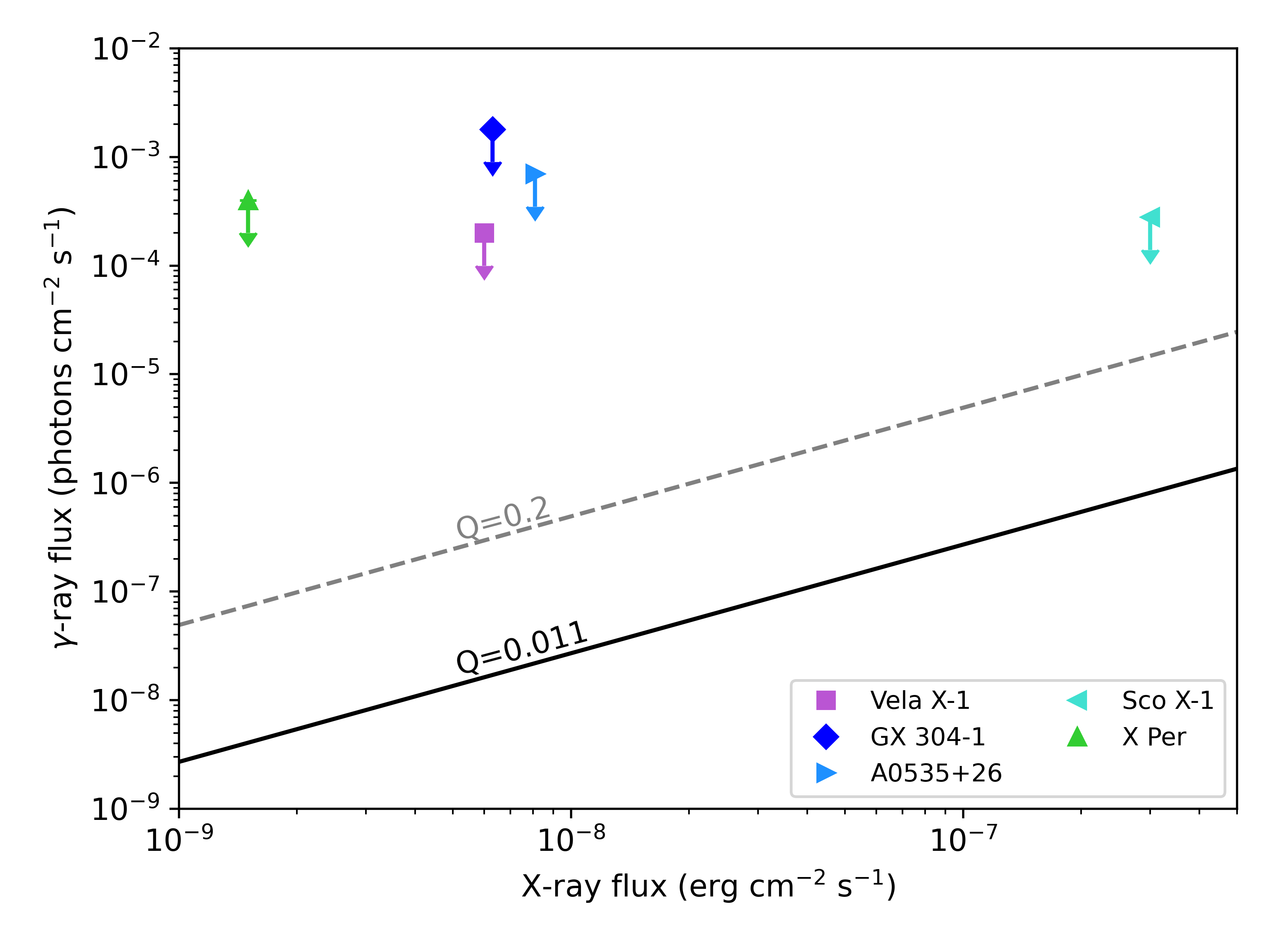}
      \caption{$3\sigma$ upper limits of the $\gamma-$ray flux in the 2.2~MeV line (assuming FWHM=10~keV) from the targets considered here, as a function of their X-ray luminosity. The solid and dashed lines show the line intensities expected from Eqs. \ref{eq:bildsten}  and \ref{eq bildsten ul2}.}
   \label{fig: all u.l.}
   \end{figure}

In Fig. \ref{fig: all u.l.} we present a comparison between the upper limits we measured and the expected emission line fluxes resulting from neutron capture on the surface of a NS.
The upper limits refer to those obtained assuming FWHM=10~keV. 
The expected redshifted 2.2~MeV line fluxes shown in this figure were obtained with Eqs. \ref{eq:bildsten} (grey line) and \ref{eq bildsten ul2} (black line) and therefore do not take the results of Sect.~\ref{sec:optimal} into account.
Nonetheless, the comparison in Fig. \ref{fig: all u.l.} is reasonable, as all the highly magnetized NSs displayed here were observed at luminosities of $L \lesssim 0.5 L_{\rm crit}$ (Sco~X-1 is the exception, likely having a low magnetic field).

%
The X-ray fluxes in Fig. \ref{fig: all u.l.} were obtained in the 0.1-100~keV range using the best-fit models derived from the SPI spectra analysed here, the JEM-X and ISGRI spectra available from the Multi-Messenger Online Data Analysis (MMODA) product gallery\footnote{\url{https://www.astro.unige.ch/mmoda/gallery/}} \citep{Neronov21}, and the spectral findings from previous studies that also analysed the data discussed in this work (refer to Sect. \ref{sect. results}).
As evident in Fig. \ref{fig: all u.l.}, the upper limits we obtained are far from the predictions.
The source whose upper limit of the $2.2$~MeV line flux is closest to the theoretical predictions is Sco~X-1.
In this regard, we point out that a lowest upper limit for Sco~X-1 exists, which was obtained with COMPTEL ($F_{2.2} \leq 2.5\times 10^{-5}$~ph~cm$^{-2}$~s$^{-1}$; \citealt{McConnell97}).

\subsection{511~keV line}

\begin{figure}
   \centering
   \includegraphics[width=\columnwidth]{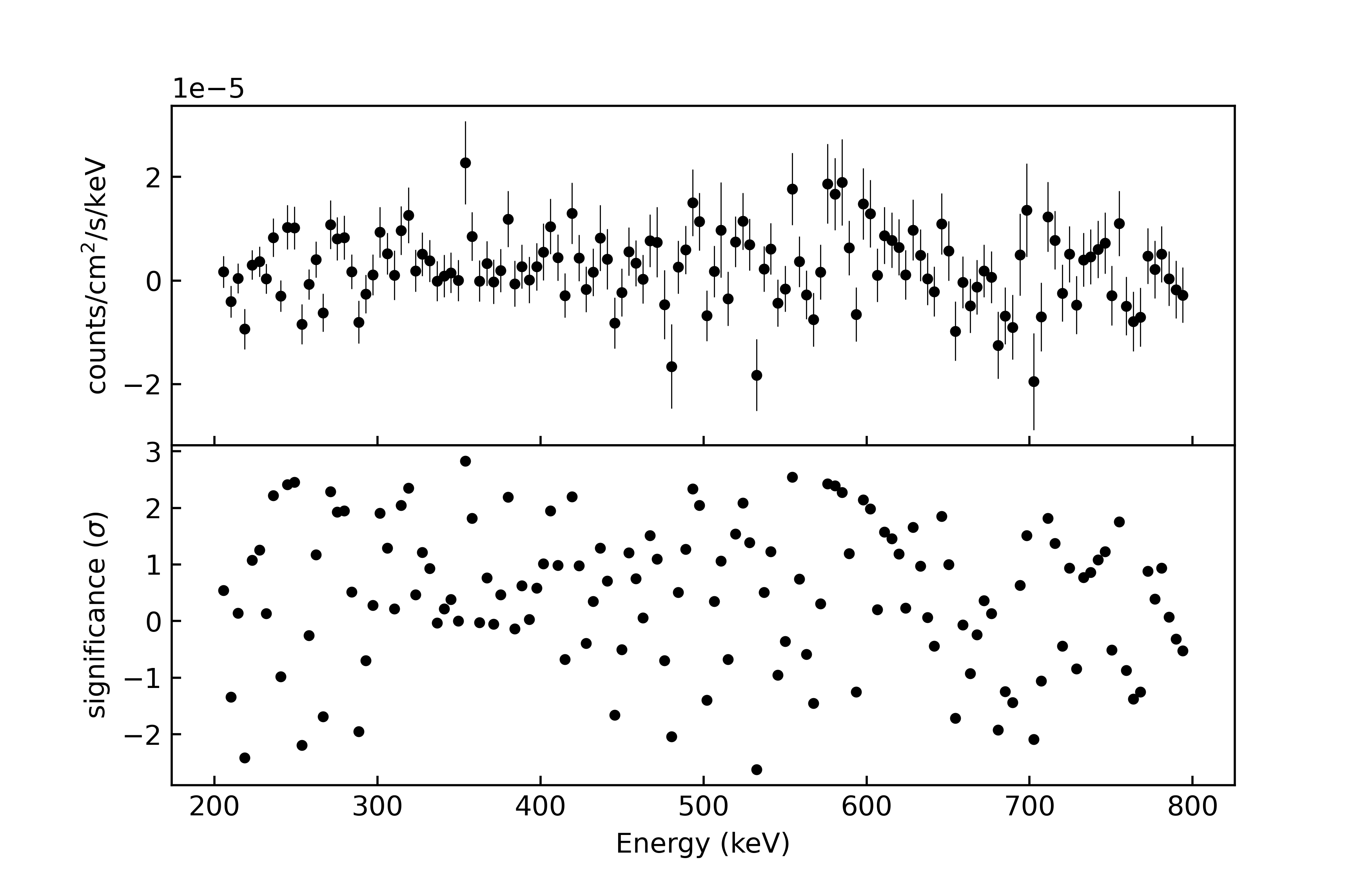}
   \caption{SPI spectrum of A~0535+26, in the energy range 200$-$800~keV. \emph{Upper panel:} Background-subtracted spectrum of A~0535+26. \emph{Lower panel:} Significance of the residuals as a function of energy. Uncertainties in these panels do not include systematic errors.}
   \label{fig: example spectrum 511}
   \end{figure}

In high magnetic field environments, $\gamma $ rays (including those from the 2.2~MeV line) with energies above the pair production threshold of $2m_{\rm e}c^2$ can generate $e^+e^-$ pairs. When the $e^+e^-$ pairs annihilate, they produce 511~keV photons, and the high-energy photon flux is reduced. \citet{Bildsten92} suggested that the 511~keV line could be detectable in accreting pulsars with $B \gtrsim 3 \times 10^{12}$ G. 
Since most of the NSs considered in this work have strong magnetic fields, with Sco~X$-$1 being the only exception, this process could produce an observable 511~keV line feature. Therefore, we searched for this using the same method used for the redshifted 2.2~MeV line, but did not detect any. The $3\sigma$ upper limits are reported in Table \ref{Table sources}. Figure \ref{fig: example spectrum 511} shows an example (A~0535+26) of one of these spectra.

\subsection{Future observations}

\begin{figure}
   \centering
   \includegraphics[width=\columnwidth]{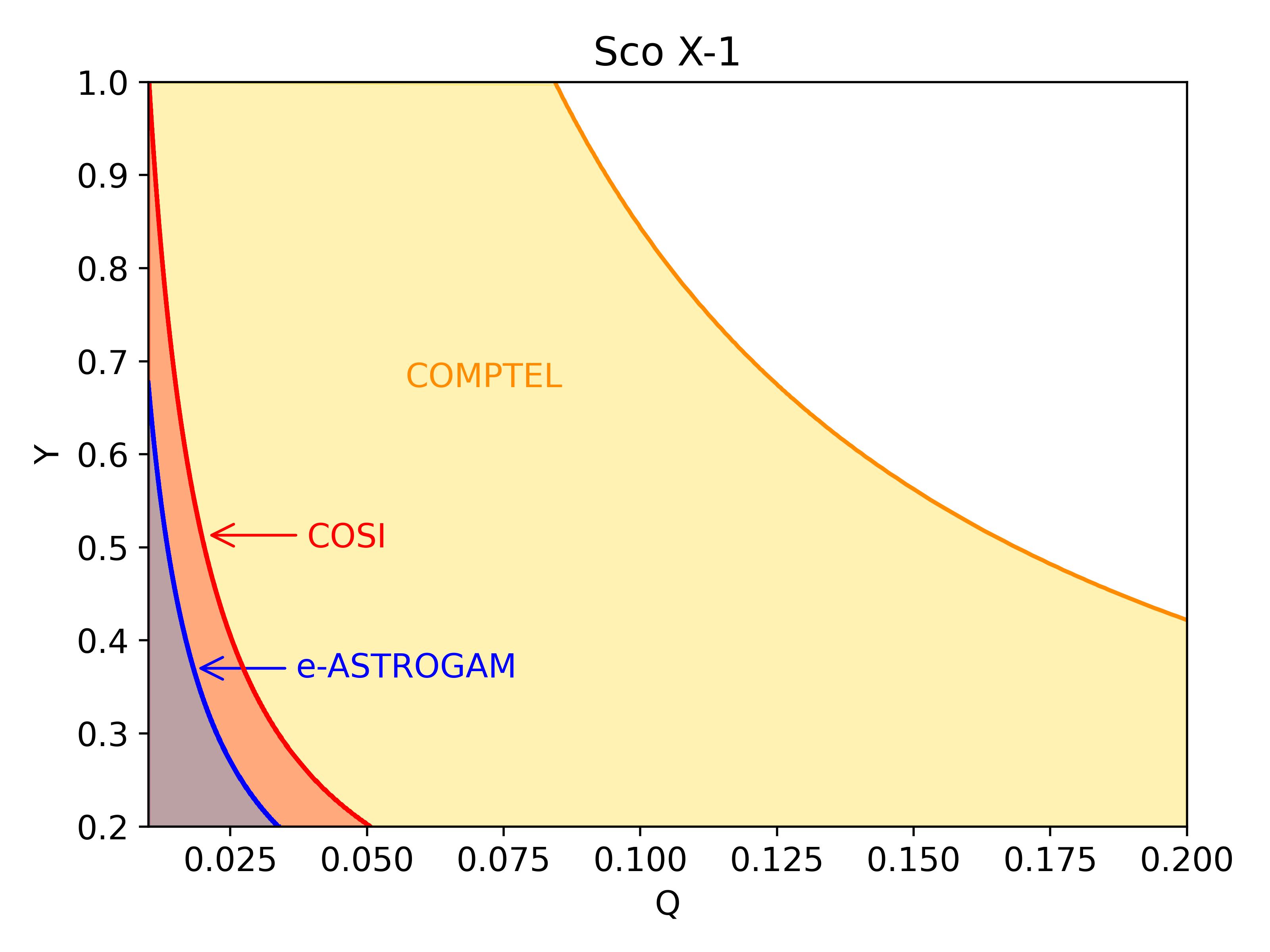}
   \caption{Comparison of achievable constraints on the parameters $Y$ and $Q$ for Sco~X-1 using Eq. \ref{eq bildsten ul2}, between the future $\gamma-$ray mission COSI, the mission concept e-ASTROGAM, and the current constraints obtainable with COMPTEL data.}
   \label{fig: comptel wow}
   \end{figure}

We point out that the MeV bandpass is currently the least explored of the whole electromagnetic spectrum (a feature often referred to as the MeV gap).
Future $\gamma-$ray spectrometers, such as the Compton Spectrometer and Imager (COSI; launch expected in 2027; \citealt{Tomsick23}), the concept missions e-ASTROGAM \citep{deAngelis18},
the MeV Astrophysical Spectroscopic Surveyor (MASS; \citealt{Zhu24}), 
the MeV Gamma-Ray Observatory (MeVGRO)\footnote{\url{https://indico.icranet.org/event/1/contributions/777/}}, and the Focusing Imager of Nuclear Astrophysics (FIONA)\footnote{\url{https://ihepbox.ihep.ac.cn/ihepbox/index.php/apps/onlyoffice/s/HCDH1xpyWM4CR7t?fileId=119688469}},
are expected to reach line sensitivity close to the theoretical predictions by the model from \citet{Bildsten93} for Sco~X-1, assuming $Q\approx 0.01$.
In particular, COSI will achieve a 3$\sigma$ line point source sensitivity in two years of survey observations of $\sim 3\times10^{-6}$~ph~cm$^{-2}$~s$^{-1}$ \citep{Tomsick23},
while e-ASTROGAM could achieve a 3$\sigma$ line point source sensitivity in $10^6$~s of $\sim 2\times10^{-6}$~ph~cm$^{-2}$~s$^{-1}$ \citep[see table 1.3.3 in: ][]{deAngelis18}.
Even if the emission line is not detected by these future instruments, the upper limits they provide could be useful in setting tighter constraints on various parameters of the theoretical models for the production of this line.
For example, the upper limits obtained with COSI and e-ASTROGAM could set limits on $Y$ and $Q$ parameters in the model from \citet{Bildsten93}.
This is shown in Fig.  \ref{fig: comptel wow}: the area subtended by the orange line shows the $Y-Q$ parameter space allowed
by the upper limit obtained with COMPTEL; similarly Fig. \ref{fig: comptel wow} shows the $Y-Q$ parameter spaces allowed by
the expected 3$\sigma$ line source sensitivity by COSI and e-ASTROGAM. SPI data does not have enough sensitivity to constrain $Q$ and $Y$.

To obtain a significant detection of the $2.2$~MeV line, a dramatic leap in the sensitivity of future MeV telescopes is needed.
Based on the current state-of-the-art technology, achieving this could be possible by using, for example, flux concentrating telescopes. These would allow for smaller detector sizes and, as such, a lower level of instrumental background.
A viable option involves the use of $\gamma-$ray optics that employ the so-called Laue lenses \citep{Frontera10}, which rely on Bragg diffraction from arrays of crystals and operate in the $\sim$MeV band (see \citealt{Virgilli22} for more on this technology and the even more advanced Fresnel lenses).
Figure 6 in \citet{Virgilli22b} shows a qualitative example of the gain in sensitivity (although in a slightly different energy band from the one of interest in this work) that could be achieved with an instrument using this type of optics.
There are also lens designs proposals aimed at extending the energy range over which Laue lenses could operate while maintaining reasonably high sensitivity. These include tunable Laue lenses \citep{Lund21a,Lund21b} and multiple layer Laue lenses \citep[see ][ and references therein]{Virgilli22}.

\section{Conclusion}

We have reported the results of searching for the redshifted line of $2.2$~MeV in a representative sample of XRBs, 
comprising both highly and low magnetized NSs.
Although we were unable to detect the line, the upper limits we obtained represent the strongest constraint ever achieved using SPI data.
From a theoretical point of view, we have shown that for highly magnetized NSs, there is an optimal X-ray luminosity to maximize the chances of detecting the 2.2~MeV line. 
It should be about half the critical luminosity; for a typical XRP this is around $\sim5\times10^{36}\,{\rm erg\,s^{-1}}$.

The upper limits obtained are higher than the expected line intensity, indicating that a significant increase in sensitivity from future MeV missions is necessary.

Finally, we note that for this study we selected a representative, though not exhaustive, sample of XRBs observed by SPI. Therefore, the search for the redshifted $2.2$~MeV line could be extended in the future to include other XRBs, exploring various luminosity states and thus fully leveraging the \inte/SPI data archive.

\begin{acknowledgements}
We thank the anonymous referee for constructive comments that helped improve the paper.
L.D. acknowledges Elisabeth Jourdain and Jean-Pierre Roques for their help on the use of {\tt spidai}. ST acknowledges support from the Alexander von Humboldt Foundation through the Friedrich Wilhelm Bessel Research Award and is grateful to the Institute for Astronomy and Astrophysics at the University of Tübingen for their hospitality.
A.M. acknowledges UKRI Stephen Hawking fellowship.
This paper is based on data from observations with INTEGRAL, an ESA project with
instruments and science data centre funded by ESA member states (especially the
PI countries: Denmark, France, Germany, Italy, Spain, and Switzerland), Czech
Republic and Poland, and with the participation of Russia and the USA.  
\end{acknowledgements}  

\bibliographystyle{aa} 
\bibliography{ld_spi}

\begin{thebibliography}{53}
\expandafter\ifx\csname natexlab\endcsname\relax\def\natexlab#1{#1}\fi

\bibitem[{{Aharonian} \& {Sunyaev}(1984)}]{Aharonian84}
{Aharonian}, F.~A. \& {Sunyaev}, R.~A. 1984, \mnras, 210, 257

\bibitem[{{Arnaud}(1996)}]{Arnaud96}
{Arnaud}, K.~A. 1996, in Astronomical Society of the Pacific Conference Series,
  Vol. 101, Astronomical Data Analysis Software and Systems V, ed. G.~H.
  {Jacoby} \& J.~{Barnes}, 17

\bibitem[{{Bailer-Jones} {et~al.}(2021){Bailer-Jones}, {Rybizki}, {Fouesneau},
  {Demleitner}, \& {Andrae}}]{Bailer21}
{Bailer-Jones}, C.~A.~L., {Rybizki}, J., {Fouesneau}, M., {Demleitner}, M., \&
  {Andrae}, R. 2021, \aj, 161, 147

\bibitem[{{Basko} \& {Sunyaev}(1976)}]{1976MNRAS.175..395B}
{Basko}, M.~M. \& {Sunyaev}, R.~A. 1976, \mnras, 175, 395

\bibitem[{{Becker} {et~al.}(2012){Becker}, {Klochkov}, {Sch{\"o}nherr},
  {Nishimura}, {Ferrigno}, {Caballero}, {Kretschmar}, {Wolff}, {Wilms}, \&
  {Staubert}}]{2012A&A...544A.123B}
{Becker}, P.~A., {Klochkov}, D., {Sch{\"o}nherr}, G., {et~al.} 2012, \aap, 544,
  A123

\bibitem[{{Bildsten} {et~al.}(1992){Bildsten}, {Salpeter}, \&
  {Wasserman}}]{Bildsten92}
{Bildsten}, L., {Salpeter}, E.~E., \& {Wasserman}, I. 1992, \apj, 384, 143

\bibitem[{{Bildsten} {et~al.}(1993){Bildsten}, {Salpeter}, \&
  {Wasserman}}]{Bildsten93}
{Bildsten}, L., {Salpeter}, E.~E., \& {Wasserman}, I. 1993, \apj, 408, 615

\bibitem[{{Boggs} \& {Smith}(2006)}]{Boggs06}
{Boggs}, S.~E. \& {Smith}, D.~M. 2006, \apjl, 637, L121

\bibitem[{{Brecher} \& {Burrows}(1980)}]{Brecher80}
{Brecher}, K. \& {Burrows}, A. 1980, \apj, 240, 642

\bibitem[{{{\c{C}}ali{\c{s}}kan} {et~al.}(2009){{\c{C}}ali{\c{s}}kan},
  {Kalemci}, {Baring}, {Boggs}, \& {Kretschmar}}]{Caliskan09}
{{\c{C}}ali{\c{s}}kan}, {\c{S}}., {Kalemci}, E., {Baring}, M.~G., {Boggs},
  S.~E., \& {Kretschmar}, P. 2009, \apj, 694, 593

\bibitem[{{Daugherty} \& {Harding}(1986)}]{1986ApJ...309..362D}
{Daugherty}, J.~K. \& {Harding}, A.~K. 1986, \apj, 309, 362

\bibitem[{{de Angelis} {et~al.}(2018){de Angelis}, {Tatischeff}, {Grenier},
  {McEnery}, {Mallamaci}, {Tavani}, {Oberlack}, {Hanlon}, {Walter}, {Argan},
  {von Ballmoos}, {Bulgarelli}, {Bykov}, {Hernanz}, {Kanbach}, {Kuvvetli},
  {Pearce}, {Zdziarski}, {Conrad}, {Ghisellini}, {Harding}, {Isern}, {Leising},
  {Longo}, {Madejski}, {Martinez}, {Mazziotta}, {Paredes}, {Pohl}, {Rando},
  {Razzano}, {Aboudan}, {Ackermann}, {Addazi}, {Ajello}, {Albertus},
  {{\'A}lvarez}, {Ambrosi}, {Ant{\'o}n}, {Antonelli}, {Babic}, {Baibussinov},
  {Balbo}, {Baldini}, {Balman}, {Bambi}, {Barres de Almeida}, {Barrio},
  {Bartels}, {Bastieri}, {Bednarek}, {Bernard}, {Bernardini}, {Bernasconi},
  {Bertucci}, {Biland}, {Bissaldi}, {Boettcher}, {Bonvicini}, {Bosch-Ramon},
  {Bottacini}, {Bozhilov}, {Bretz}, {Branchesi}, {Brdar}, {Bringmann},
  {Brogna}, {Budtz J{\o}rgensen}, {Busetto}, {Buson}, {Busso}, {Caccianiga},
  {Camera}, {Campana}, {Caraveo}, {Cardillo}, {Carlson}, {Celestin},
  {Cerme{\~n}o}, {Chen}, {Cheung}, {Churazov}, {Ciprini}, {Coc},
  {Colafrancesco}, {Coleiro}, {Collmar}, {Coppi}, {Curado da Silva}, {Cutini},
  {D'Ammando}, {de Lotto}, {de Martino}, {De Rosa}, {Del Santo}, {Delgado},
  {Diehl}, {Dietrich}, {Dolgov}, {Dom{\'\i}nguez}, {Dominis Prester},
  {Donnarumma}, {Dorner}, {Doro}, {Dutra}, {Elsaesser}, {Fabrizio},
  {Fern{\'a}ndez-Barral}, {Fioretti}, {Foffano}, {Formato}, {Fornengo},
  {Foschini}, {Franceschini}, {Franckowiak}, {Funk}, {Fuschino}, {Gaggero},
  {Galanti}, {Gargano}, {Gasparrini}, {Gehrz}, {Giammaria}, {Giglietto},
  {Giommi}, {Giordano}, {Giroletti}, {Ghirlanda}, {Godinovic}, {Gouiff{\'e}s},
  {Grove}, {Hamadache}, {Hartmann}, {Hayashida}, {Hryczuk}, {Jean}, {Johnson},
  {Jos{\'e}}, {Kaufmann}, {Khelifi}, {Kiener}, {Kn{\"o}dlseder}, {Kole},
  {Kopp}, {Kozhuharov}, {Labanti}, {Lalkovski}, {Laurent}, {Limousin},
  {Linares}, {Lindfors}, {Lindner}, {Liu}, {Lombardi}, {Loparco},
  {L{\'o}pez-Coto}, {L{\'o}pez Moya}, {Lott}, {Lubrano}, {Malyshev},
  {Mankuzhiyil}, {Mannheim}, {March{\~a}}, {Marcian{\`o}}, {Marcote},
  {Mariotti}, {Marisaldi}, {McBreen}, {Mereghetti}, {Merle}, {Mignani},
  {Minervini}, {Moiseev}, {Morselli}, {Moura}, {Nakazawa}, {Nava}, {Nieto},
  {Orienti}, {Orio}, {Orlando}, {Orleanski}, {Paiano}, {Paoletti}, {Papitto},
  {Pasquato}, {Patricelli}, {P{\'e}rez-Garc{\'\i}a}, {Persic}, {Piano},
  {Pichel}, {Pimenta}, {Pittori}, {Porter}, {Poutanen}, {Prandini}, {Prantzos},
  {Produit}, {Profumo}, {Queiroz}, {Rain{\'o}}, {Raklev}, {Regis}, {Reichardt},
  {Rephaeli}, {Rico}, {Rodejohann}, {Rodriguez Fernandez}, {Roncadelli},
  {Roso}, {Rovero}, {Ruffini}, {Sala}, {S{\'a}nchez-Conde}, {Santangelo}, {Saz
  Parkinson}, {Sbarrato}, {Shearer}, {Shellard}, {Short}, {Siegert},
  {Siqueira}, {Spinelli}, {Stamerra}, {Starrfield}, {Strong}, {Str{\"u}mke},
  {Tavecchio}, {Taverna}, {Terzi{\'c}}, {Thompson}, {Tibolla}, {Torres},
  {Turolla}, {Ulyanov}, {Ursi}, {Vacchi}, {van den Abeele},
  {Vankova-Kirilovai}, {Venter}, {Verrecchia}, {Vincent}, {Wang}, {Weniger},
  {Wu}, {Zaharija{\v{s}}}, {Zampieri}, {Zane}, {Zimmer}, {Zoglauer}, \&
  {E-Astrogam Collaboration}}]{deAngelis18}
{de Angelis}, A., {Tatischeff}, V., {Grenier}, I.~A., {et~al.} 2018, Journal of
  High Energy Astrophysics, 19, 1

\bibitem[{{Diehl} {et~al.}(2018){Diehl}, {Siegert}, {Greiner}, {Krause},
  {Kretschmer}, {Lang}, {Pleintinger}, {Strong}, {Weinberger}, \&
  {Zhang}}]{Diehl18}
{Diehl}, R., {Siegert}, T., {Greiner}, J., {et~al.} 2018, \aap, 611, A12

\bibitem[{{Falanga} {et~al.}(2015){Falanga}, {Bozzo}, {Lutovinov},
  {Bonnet-Bidaud}, {Fetisova}, \& {Puls}}]{Falanga15}
{Falanga}, M., {Bozzo}, E., {Lutovinov}, A., {et~al.} 2015, \aap, 577, A130

\bibitem[{{Ferrigno} {et~al.}(2016){Ferrigno}, {Ducci}, {Bozzo}, {Kretschmar},
  {K{\"u}hnel}, {Malacaria}, {Pottschmidt}, {Santangelo}, {Savchenko}, \&
  {Wilms}}]{Ferrigno16}
{Ferrigno}, C., {Ducci}, L., {Bozzo}, E., {et~al.} 2016, \aap, 595, A17

\bibitem[{{Frontera} \& {von Ballmoos}(2010)}]{Frontera10}
{Frontera}, F. \& {von Ballmoos}, P. 2010, X-Ray Optics and Instrumentation
  2010, 2010, 215375

\bibitem[{{Galaudage} {et~al.}(2022){Galaudage}, {Wette}, {Galloway}, \&
  {Messenger}}]{Galaudage22}
{Galaudage}, S., {Wette}, K., {Galloway}, D.~K., \& {Messenger}, C. 2022,
  \mnras, 509, 1745

\bibitem[{{Guessoum} \& {Jean}(2002)}]{Guessoum02}
{Guessoum}, N. \& {Jean}, P. 2002, \aap, 396, 157

\bibitem[{{Guessoum} \& {Jean}(2004)}]{Guessoum04}
{Guessoum}, N. \& {Jean}, P. 2004, Nuclear Physics B Proceedings Supplements,
  132, 396

\bibitem[{{Harris} \& {Share}(1991)}]{Harris91}
{Harris}, M.~J. \& {Share}, G.~H. 1991, \apj, 381, 439

\bibitem[{{Jean} \& {Guessoum}(2001)}]{Jean01}
{Jean}, P. \& {Guessoum}, N. 2001, \aap, 378, 509

\bibitem[{{Jourdain} \& {Roques}(2009)}]{Jourdain09}
{Jourdain}, E. \& {Roques}, J.~P. 2009, \apj, 704, 17

\bibitem[{{Klochkov} {et~al.}(2012){Klochkov}, {Doroshenko}, {Santangelo},
  {Staubert}, {Ferrigno}, {Kretschmar}, {Caballero}, {Wilms}, {Kreykenbohm},
  {Pottschmidt}, {Rothschild}, {Wilson-Hodge}, \& {P{\"u}hlhofer}}]{Klochkov12}
{Klochkov}, D., {Doroshenko}, V., {Santangelo}, A., {et~al.} 2012, \aap, 542,
  L28

\bibitem[{{Lund}(2021{\natexlab{a}})}]{Lund21a}
{Lund}, N. 2021{\natexlab{a}}, Experimental Astronomy, 51, 165

\bibitem[{{Lund}(2021{\natexlab{b}})}]{Lund21b}
{Lund}, N. 2021{\natexlab{b}}, Experimental Astronomy, 51, 153

\bibitem[{{Malacaria} {et~al.}(2015){Malacaria}, {Klochkov}, {Santangelo}, \&
  {Staubert}}]{Malacaria15}
{Malacaria}, C., {Klochkov}, D., {Santangelo}, A., \& {Staubert}, R. 2015,
  \aap, 581, A121

\bibitem[{{Markozov} {et~al.}(2023){Markozov}, {Kaminker}, \&
  {Potekhin}}]{2023AstL...49..583M}
{Markozov}, I.~D., {Kaminker}, A.~D., \& {Potekhin}, A.~Y. 2023, Astronomy
  Letters, 49, 583

\bibitem[{{Markozov} \& {Mushtukov}(2024)}]{2024MNRAS.527.5374M}
{Markozov}, I.~D. \& {Mushtukov}, A.~A. 2024, \mnras, 527, 5374

\bibitem[{{McConnell} {et~al.}(1997){McConnell}, {Fletcher}, {Bennett},
  {Bloemen}, {Diehl}, {Hermsen}, {Ryan}, {Sch{\"o}nfelder}, {Strong}, \& {van
  Dijk}}]{McConnell97}
{McConnell}, M., {Fletcher}, S., {Bennett}, K., {et~al.} 1997, in American
  Institute of Physics Conference Series, Vol. 410, Proceedings of the Fourth
  Compton Symposium, ed. C.~D. {Dermer}, M.~S. {Strickman}, \& J.~D. {Kurfess}
  (AIP), 1099--1103

\bibitem[{{Mighell}(1999)}]{Mighell99}
{Mighell}, K.~J. 1999, \apj, 518, 380

\bibitem[{{Mushtukov} \& {Tsygankov}(2022)}]{2022arXiv220414185M}
{Mushtukov}, A. \& {Tsygankov}, S. 2022, arXiv e-prints, arXiv:2204.14185

\bibitem[{{Mushtukov} {et~al.}(2015{\natexlab{a}}){Mushtukov}, {Suleimanov},
  {Tsygankov}, \& {Poutanen}}]{2015MNRAS.447.1847M}
{Mushtukov}, A.~A., {Suleimanov}, V.~F., {Tsygankov}, S.~S., \& {Poutanen}, J.
  2015{\natexlab{a}}, \mnras, 447, 1847

\bibitem[{{Mushtukov} {et~al.}(2015{\natexlab{b}}){Mushtukov}, {Tsygankov},
  {Serber}, {Suleimanov}, \& {Poutanen}}]{2015MNRAS.454.2714M}
{Mushtukov}, A.~A., {Tsygankov}, S.~S., {Serber}, A.~V., {Suleimanov}, V.~F.,
  \& {Poutanen}, J. 2015{\natexlab{b}}, \mnras, 454, 2714

\bibitem[{{Neronov} {et~al.}(2021){Neronov}, {Savchenko}, {Tramacere},
  {Meharga}, {Ferrigno}, \& {Paltani}}]{Neronov21}
{Neronov}, A., {Savchenko}, V., {Tramacere}, A., {et~al.} 2021, \aap, 651, A97

\bibitem[{{{\"O}zel} \& {Psaltis}(2003)}]{Ozel03}
{{\"O}zel}, F. \& {Psaltis}, D. 2003, \apjl, 582, L31

\bibitem[{{Reig} \& {Roche}(1999)}]{Reig99}
{Reig}, P. \& {Roche}, P. 1999, \mnras, 306, 100

\bibitem[{{Reina} {et~al.}(1974){Reina}, {Treves}, \& {Tarenghi}}]{Reina74}
{Reina}, C., {Treves}, A., \& {Tarenghi}, M. 1974, \aap, 32, 317

\bibitem[{{Roques} \& {Jourdain}(2019)}]{Roques19}
{Roques}, J.-P. \& {Jourdain}, E. 2019, \apj, 870, 92

\bibitem[{{Roques} {et~al.}(2003){Roques}, {Schanne}, {von Kienlin},
  {Kn{\"o}dlseder}, {Briet}, {Bouchet}, {Paul}, {Boggs}, {Caraveo},
  {Cass{\'e}}, {Cordier}, {Diehl}, {Durouchoux}, {Jean}, {Leleux}, {Lichti},
  {Mandrou}, {Matteson}, {Sanchez}, {Sch{\"o}nfelder}, {Skinner}, {Strong},
  {Teegarden}, {Vedrenne}, {von Ballmoos}, \& {Wunderer}}]{Roques03}
{Roques}, J.~P., {Schanne}, S., {von Kienlin}, A., {et~al.} 2003, \aap, 411,
  L91

\bibitem[{{Rosenberg} {et~al.}(1975){Rosenberg}, {Eyles}, {Skinner}, \&
  {Willmore}}]{Rosenberg75}
{Rosenberg}, F.~D., {Eyles}, C.~J., {Skinner}, G.~K., \& {Willmore}, A.~P.
  1975, \nat, 256, 628

\bibitem[{{Salganik} {et~al.}(2023){Salganik}, {Tsygankov}, {Doroshenko},
  {Molkov}, {Lutovinov}, {Mushtukov}, \& {Poutanen}}]{Salganik23}
{Salganik}, A., {Tsygankov}, S.~S., {Doroshenko}, V., {et~al.} 2023, \mnras,
  524, 5213

\bibitem[{{Sartore} {et~al.}(2015){Sartore}, {Jourdain}, \&
  {Roques}}]{Sartore15}
{Sartore}, N., {Jourdain}, E., \& {Roques}, J.~P. 2015, \apj, 806, 193

\bibitem[{{Shvartsman}(1970)}]{Shvartsman70}
{Shvartsman}, V.~F. 1970, Astrophysics, 6, 56

\bibitem[{Siemiginowska(2011)}]{Siemiginowska11}
Siemiginowska, A. 2011, Statistics, ed. K.~Arnaud, R.~Smith, \&
  A.~Siemiginowska, Cambridge Observing Handbooks for Research Astronomers
  (Cambridge University Press), 131–145

\bibitem[{{Teegarden} \& {Watanabe}(2006)}]{Teegarden06}
{Teegarden}, B.~J. \& {Watanabe}, K. 2006, \apj, 646, 965

\bibitem[{{Tomsick} {et~al.}(2023){Tomsick}, {Boggs}, {Zoglauer}, {Hartmann},
  {Ajello}, {Burns}, {Fryer}, {Karwin}, {Kierans}, {Lowell}, {Malzac},
  {Roberts}, {Saint-Hilaire}, {Shih}, {Siegert}, {Sleator}, {Takahashi},
  {Tavecchio}, {Wulf}, {Beechert}, {Gulick}, {Joens}, {Lazar}, {Neights},
  {Martinez Oliveros}, {Matsumoto}, {Melia}, {Yoneda}, {Amman}, {Bal}, {von
  Ballmoos}, {Bates}, {B{\"o}ttcher}, {Bulgarelli}, {Cavazzuti}, {Chang},
  {Chen}, {Chu}, {Ciabattoni}, {Costamante}, {Dreyer}, {Fioretti}, {Fenu},
  {Gallego}, {Ghirlanda}, {Grove}, {Huang}, {Jean}, {Khatiya},
  {Kn{\"o}dlseder}, {Krause}, {Leising}, {Lewis}, {Lommler}, {Marcotulli},
  {Martinez-Castellanos}, {Mittal}, {Negro}, {Al Nussirat}, {Nakazawa},
  {Oberlack}, {Palmore}, {Panebianco}, {Parmiggiani}, {Parsotan}, {Pike},
  {Rogers}, {Schutte}, {Sheng}, {Smale}, {Smith}, {Trigg}, {Venters},
  {Watanabe}, \& {Zhang}}]{Tomsick23}
{Tomsick}, J.~A., {Boggs}, S.~E., {Zoglauer}, A., {et~al.} 2023, arXiv
  e-prints, arXiv:2308.12362

\bibitem[{{Vedrenne} {et~al.}(2003){Vedrenne}, {Roques}, {Sch{\"o}nfelder},
  {Mandrou}, {Lichti}, {von Kienlin}, {Cordier}, {Schanne}, {Kn{\"o}dlseder},
  {Skinner}, {Jean}, {Sanchez}, {Caraveo}, {Teegarden}, {von Ballmoos},
  {Bouchet}, {Paul}, {Matteson}, {Boggs}, {Wunderer}, {Leleux},
  {Weidenspointner}, {Durouchoux}, {Diehl}, {Strong}, {Cass{\'e}}, {Clair}, \&
  {Andr{\'e}}}]{Vedrenne03}
{Vedrenne}, G., {Roques}, J.~P., {Sch{\"o}nfelder}, V., {et~al.} 2003, \aap,
  411, L63

\bibitem[{{Vestrand}(1989)}]{Vestrand89}
{Vestrand}, W. 1989, in Proceedings of the Gamma-Ray Observatory Workshop, ed.
  N.~Johnson, 4--274

\bibitem[{{Virgilli} {et~al.}(2022{\natexlab{a}}){Virgilli}, {Frontera},
  {Rosati}, {Guidorzi}, {Ferro}, {Moita}, {Orlandini}, {Fuschino}, {Campana},
  {Labanti}, {Marchesini}, {Caroli}, {Auricchio}, {Stephen}, {Ferrari},
  {Squerzanti}, {Del Sordo}, {Gargano}, \& {Pucci}}]{Virgilli22b}
{Virgilli}, E., {Frontera}, F., {Rosati}, P., {et~al.} 2022{\natexlab{a}},
  arXiv e-prints, arXiv:2211.16916

\bibitem[{{Virgilli} {et~al.}(2022{\natexlab{b}}){Virgilli}, {Halloin}, \&
  {Skinner}}]{Virgilli22}
{Virgilli}, E., {Halloin}, H., \& {Skinner}, G. 2022{\natexlab{b}}, Laue and
  Fresnel Lenses, ed. C.~Bambi \& A.~Santangelo (Singapore: Springer Nature
  Singapore), 1--39

\bibitem[{{Vrtilek} {et~al.}(1991){Vrtilek}, {Penninx}, {Raymond}, {Verbunt},
  {Hertz}, {Wood}, {Lewin}, \& {Mitsuda}}]{Vrtilek91}
{Vrtilek}, S.~D., {Penninx}, W., {Raymond}, J.~C., {et~al.} 1991, \apj, 376,
  278

\bibitem[{{White} {et~al.}(1976){White}, {Mason}, {Sanford}, \&
  {Murdin}}]{White76}
{White}, N.~E., {Mason}, K.~O., {Sanford}, P.~W., \& {Murdin}, P. 1976, \mnras,
  176, 201

\bibitem[{{Zhu} {et~al.}(2024){Zhu}, {Zheng}, {Feng}, {Zeng}, {Huang},
  {Hsiang-Yue}, {Chang}, {Li}, {Chang}, {Pan}, {Ma}, {Wu}, {Li}, {Bai}, {Ge},
  {Ji}, {Li}, {Shen}, {Wang}, {Wang}, {Zhang}, \& {Zhang}}]{Zhu24}
{Zhu}, J., {Zheng}, X., {Feng}, H., {et~al.} 2024, Experimental Astronomy, 57,
  2

\end{thebibliography}

\begin{appendix}
\section{SPI upper limits in highly magnetized neutron stars during outbursts}

In Sect. \ref{sect. data analysis} we selected the SPI data to search for the redshifted 2.2~MeV line based on the theoretical results presented in Sect. \ref{sec:optimal}. Specifically, we focused on observations of highly magnetized NSs with luminosities $\lesssim 0.5L_{\rm crit}$. One might question whether this emission line could be detected in these sources at higher luminosity states, regardless of the theoretical evidence discussed in Sect. \ref{sec:optimal}. Although we consider this hypothesis unlikely, for the sake of completeness, we present in this appendix the upper limits obtained from the search for the redshifted 2.2~MeV line in A~0535+26, GX~304$-$1, RX~J0440.9+4431, and Swift~J0243.6+6124 during bright outbursts detected with SPI.
These observations are summarized in Table \ref{table:log2}.

\begin{table}
\begin{center}
\caption{Time intervals of the observations analysed in this work.}
\vspace{-0.3cm}
\label{table:log2}
\resizebox{\columnwidth}{!}{
\begin{tabular}{lcc}
\hline
\hline
\noalign{\smallskip}
Source name         & time interval &  \inte\ revolution$^a$   \\
\noalign{\smallskip}
\hline
\noalign{\smallskip}
A~0535+26            & 16 February - 11 March 2011 & 1019-1026 \\
\noalign{\smallskip}
\hline
\noalign{\smallskip}
V~0332+53           & 8 January - 21 February 2005 & 273-288 \\
                    & 17 July 2015 - 7 October 2015 & 1565, 1570. 1596 \\
\noalign{\smallskip}
\hline
\noalign{\smallskip}
GX~304$-$1           & 20 January - 1 February 2012 & 1132-1136 \\
\noalign{\smallskip}
\hline
\noalign{\smallskip}
RX~J0440.9+4431      & 28 January - 4 February 2023    & 2600-2602 \\
\noalign{\smallskip}
\hline
\noalign{\smallskip}
Swift~J0243.6+6124   &  8 - 30 July 2023                & 2660-2668\\ 
\noalign{\smallskip}
\hline 
\end{tabular}
}
\end{center}
 {\small Notes. \\
 \noindent $^a$ The numbers of \inte\ revolutions around the Earth (starting from 0 at launch) are commonly used to refer to the corresponding observations.\\
  }
\end{table}

For A~0535+26, we considered the giant outburst that occurred in 2011 \citep{Sartore15}. 
For V~0332+53, we considered both giant outbursts observed by \inte\ in 2004-2005 and in 2015 \citep{Ferrigno16}.
For GX~304$-$1, we considered the bright outburst of 2012 \citep{Klochkov12, Malacaria15}.
For RX~J0440.9+4431, we considered the outburst observed by \inte\ in 2023 \citep{Salganik23}. 
For Swift~J0243.6+6124, we considered the outburst observed by \inte\ in 2023. 

Table \ref{Table a} shows, for each source, the net exposure time and the $3\sigma$ upper limits on the redshifted 2.2~MeV and 511~keV lines.

\begin{table*}
\begin{center}
\caption{Highly magnetized accreting NSs during bright X-ray luminosity states, along with their respective net exposure times for the SPI data, and the highest $3\sigma$ upper limits in the ranges 1$-$2.2~MeV (for a redshifted 2.2~MeV line) and 200-800~keV (for a 511~keV line), assuming different values of FWHM.}
\vspace{-0.3cm}
\label{Table a}
\resizebox{\columnwidth+\columnwidth}{!}{
\begin{tabular}{lcccccccccccc}
\hline
\hline
\noalign{\smallskip}
Source name         & net exposure &  $L_{\rm x}$$^a$ & distance$^b$         & \multicolumn{5}{c}{$3\sigma$ u.l. 2.2~MeV line}  &  \multicolumn{4}{c}{$3\sigma$ u.l. 511~keV line} \\
\noalign{\smallskip}
                    &   (ks) &  (erg~s$^{-1}$)   &  (kpc)              & \multicolumn{5}{c}{$10^{-4}$ph~cm$^{-2}$~s$^{-1}$}  &   \multicolumn{4}{c}{$10^{-3}$ph~cm$^{-2}$~s$^{-1}$}     \\
                    &        &                   &                     &FWHM:&    10   &    20   &    40   &    100    &FWHM:&    10   &    20     &     40    \\
                    &        &                   &                     &(keV)&         &         &         &           &(keV)&         &           &           \\
\noalign{\smallskip}
\hline
\noalign{\smallskip}
A~0535+26            & 393   &  $2.4\times 10^{37}$ & $1.8\pm0.1$      &            & 1.2  &  1.4  &  1.5  &  2.4  & &   1.5  &  1.4  &  1.8  \\
\noalign{\smallskip}
V~0332+53            & 596   &  $3\times 10^{37}$   & $5.6\pm0.7$      &           &  5.8  &  6.5  & 8.5  &   8.6  & & 0.5   & 0.7  &  0.9     \\
\noalign{\smallskip}
GX~304$-$1          & 122    &  $6.2\times 10^{36}$ & $1.86\pm0.04$    &           &  17   &  23   &  32   &  45   & &  2.4    & 3.1  &  3.2    \\
\noalign{\smallskip}
Swift~J0243.6+6124  & 159    &  $9.8\times 10^{37}$ & $5.2\pm0.3$      &           &  11   &  12   &  16   &  22   &  & 1.8    &  2.1  &  2.6     \\
\noalign{\smallskip}
RX~J0440.9+4431     & 255    &  $5\times 10^{37}$ & $2.44{+0.06\atop -0.08}$ &      &  11  &  15   &  17   &  32    & & 1.7   &  2.0   &  2.4    \\
\noalign{\smallskip}
\hline 
\end{tabular}
}
\end{center}
 {\small Notes. \\
 \noindent $^a$ X-ray luminosity, in the energy range 0.1--100~keV.\\
 \noindent $^b$ distances from \citet{Bailer21}.\\
 }
\end{table*}

\section{200-800~keV and 1-2.2~MeV upper limits} \label{app. B}

Table \ref{table B} shows the $3\sigma$ upper limits of the fluxes in the energy bands 200-800 keV and 1-2.2~MeV. For these calculations, we extracted the spectra in the energy range of 20-2200~keV adopting broader energy bins than those used in Sect. \ref{sect. data analysis}, 15 bins in log-scale. We then fitted the spectra above 200~keV with a power law using the spectral model {\tt pegpwrlw} in {\tt xspec}.

\begin{table}
\begin{center}
\caption{Accreting NSs analysed in this work, along with their respective $3\sigma$ upper limits in the energy bands 200-800~keV and 1$-$2.2~MeV.}
\vspace{-0.3cm}
\label{table B}
\begin{tabular}{lcc}
\hline
\hline
\noalign{\smallskip}
Source name         &   200-800~keV$^a$       &       1-2.2~MeV$^a$    \\
\noalign{\smallskip}
\hline
\noalign{\smallskip}
A~0535+26            &         1.1            &          4.6           \\
\noalign{\smallskip}
GX~304$-$1          &          0.5            &          4.3           \\ 
\noalign{\smallskip}
Vela X-1          &            0.13           &          0.16           \\
\noalign{\smallskip}
X~Persei          &            0.7            &          1.5            \\
\noalign{\smallskip}
Sco X-1           &            0.3            &          2.8            \\
\noalign{\smallskip}
Swift~J0243.6+6124  &          0.9            &          4.5            \\
\noalign{\smallskip}
RX~J0440.9+4431     &          2.4            &          3.3            \\
\noalign{\smallskip}
\hline 
\end{tabular}
\end{center}
 {\small Notes. \\
 \noindent $^a$ in units of $10^{-9}$~erg~cm$^{-2}$~s$^{-1}$ 
 }
\end{table}

 \end{appendix}

\end{document}